\documentclass[runningheads,a4paper,oribibl]{llncs}

\usepackage{amssymb}
\usepackage[dvips,final,pdftex]{graphicx}
\usepackage[utf8]{inputenc}
\usepackage{url}
\usepackage{verbatim}
\usepackage{booktabs}

\urldef{\mailsa}\path|{pedro.chahuara, lampert, pierre.gancarski}@unistra.fr|  
\newcommand{\keywords}[1]{\par\addvspace\baselineskip
\noindent\keywordname\enspace\ignorespaces#1}

\usepackage{color}
\usepackage{siunitx}
\usepackage{epsfig}
\usepackage{psfrag}
\usepackage[crop=pdfcrop]{pstool}
\usepackage{multirow}
\usepackage{amsmath}
\usepackage{cite}
\usepackage{enumitem}

\DeclareMathOperator{\Multi}{Mult}

\DeclareMathOperator{\similarity}{Similarity}

\begin{document}

\mainmatter  

\title{Retrieving and Ranking Similar Questions from Question-Answer Archives Using Topic Modelling and Topic Distribution Regression}

\titlerunning{Retrieving and Ranking Similar Questions from Question-Answer Archives}

%
%
\author{Pedro Chahuara\inst{1,2}\and Thomas Lampert\inst{1,3}\and Pierre Gan\c{c}arski\inst{1}}

\authorrunning{P.\ Chahuara, T.\ Lampert, and P.\ Gan\c{c}arski}

\institute{Laboratoire ICube, Universit\'{e} de Strasbourg, France\\
\mailsa\\
\and
Xerox Research Centre Europe, Meylan, France
\and
Laboratoire Quantup, Strasbourg, France
}

%
%

\toctitle{Retrieving and Ranking Similar Questions from Question-Answer Archives Using Topic Modelling and Topic Distribution Regression}
\tocauthor{Pedro Chahuara et al.}
\maketitle

\setcounter{footnote}{0}

\begin{abstract}
Presented herein is a novel model for similar question ranking within collaborative question answer platforms. The presented approach integrates a regression stage to relate topics derived from questions to those derived from question-answer pairs. This helps to avoid problems caused by the differences in vocabulary used within questions and answers, and the tendency for questions to be shorter than answers. The performance of the model is shown to outperform translation methods and topic modelling (without regression) on several real-world datasets.
\end{abstract}

\section{Introduction}

During the last decade internet based Collaborative Question Answering (CQA) platforms have increased in popularity. These platforms offer a social environment for people to seek answers to questions, and where the answers are offered by other community members. Users pose questions in natural language, as opposed to queries in web search engines, and community members propose answers in addition to voting and rating the information posted on the platform. Some of the most popular CQA sites are Yahoo!\ Questions, Quora, and StackExchange. Besides public CQA websites, similar systems can be found in industry, for example in retail and business websites where users can pose questions about a company's product and a group of specialists can give support. 

This content has attracted the attention of researchers from a number of domains \cite{Jeon2005,zhang2014,Wang09,Cui05,xue2008,Lee2008,Bernhard2009} who aim to automatically return existing, relevant information from the CQA database when a novel question is submitted. Proposed approaches fall into two categories: determining the most relevant answers to a question \cite{Yang2013, Berger00}; and determining similar questions \cite{Yang2013,Jeon2005,xue2008}. The latter is the problem that is covered in the present work. As such, the system helps to remove the delay needed for other community members to answer; and the list of related questions provides material for users to acquire more knowledge on the topic of their question.

Solving this problem is not a trivial matter as semantically similar questions and answers can be lexically dissimilar \cite{zhang2014, Jeon2005}, referred to as the `lexical chasm' \cite{Berger00}. For instance the questions ``Where can I watch movies on the internet for free?'' and ``Are there any sites for streaming films?'' are semantically related but lexically different. The opposite case is also possible---questions having words in common may have different semantic meanings. Besides the need for accurately identifying a question's semantics, a solution to the problem must deal with noisy information such as: misspelt words, polysemy, and short questions.

Similar questions are typically found by comparing the query question to the content of existing questions as it has been shown that finding similar questions based solely on their answers does not perform well \cite{Jeon2005, xue2008}. Nevertheless Xue et al.\ demonstrated that combining information derived from existing questions and their answers outperforms the other strategies \cite{xue2008}. In recent years topic modelling has been applied to this problem \cite{zhang2014, Cai11, Vasiljevic16} as it reduces the dimensionality of textual information when compared to classical methods such as bag-of words and efficiently handles polysemy and synonymy. These approaches, however, have thus far only been used to model the questions in the archives. As such, the contribution of the present work is twofold: firstly, the application of Latent Dirichlet Allocation (LDA) to model topics among the questions and answers in the archive; and secondly, the use of a regression step to estimate the appropriate QA topic distribution from that of a novel question.

This paper is organized as follows: Section \ref{sec:relatedwork} presents the state of the art, Section \ref{sec:methodolgy} the study's methodology, Section \ref{sec:experiment} the experimental setup and results, which are discussed in Section \ref{sec:discussion}, and conclusions are presented in Section \ref{sec:conclusions}.

\section{Related Work}
\label{sec:relatedwork}

The principal challenge when retrieving related questions and answers in a QA database given a new question is the lexical gap that may exist between two semantically similar questions. In general, a method that intends to solve the problem of question retrieval should be composed at least of two main parts: a document representation that can properly express the semantics and context of QAs in the database; and a mechanism for comparing the similarity of documents given their representations. The most widespread document representation methods in the literature are those based on bag-of-words (BOW), which explicitly represents each of the document's words. Comparison is achieved by computing the number of matching words between two BOW representations. There exist several variations of this class of methods, each weighting words that have specific properties in the dataset, such as tf-idf and BM25 \cite{Robertson1994}. This class of methods is able to measure two documents' lexical similarity but it does not capture information regarding their semantics and context.

In QA databases, questions and answers are often short and contain many word variations resulting from grammatical inflection, misspelling, and informal abbreviations. As a consequence, BOW representations in QA corpora produce a vector representation that can be too sparse. Besides sparsity, BOW representations do not provide a measure of co-occurrence or shared contextual information, which can increase the similarity of related documents.

An approach that overcomes these limitations is the translation model, first proposed for use in this context by Jeon et al.\ \cite{Jeon2005}. Their method consists of two stages: first a set of semantically similar questions are found by matching their answers using a query-likelihood language model; and subsequently, word translation probabilities are estimated using the IBM translation model 1 \cite{Brown93}. Several extensions have been proposed \cite{xue2008,Lee2008,Bernhard2009} including the use of external corpora \cite{Zhou2013,singh2012} such as Wikipedia. Xue et al.\ \cite{xue2008} propose an extension that combines the IBM translation model (applied to the questions) with a query likelihood language model (applied to the answers). Translation-based models have become the state-of-the-art in query retrieval \cite{Cai11, Zhou2013a} but they suffer from some limitations: they do not capture word co-occurrences nor word distributions in the corpora.

In the last decade Topic Modeling has become an important method for text analysis. Since the topics that characterise a document can be considered a semantic representation, it is possible to use topic distributions inferred using a method such as Latent Dirichlet Allocation (LDA) \cite{Blei2003} to measure the semantic similarity between documents in a corpora. Consequently, several approaches for applying topic modelling to QA archives have been proposed: Zhang et al.\ \cite{zhang2014} retrieve similar questions by measuring lexical and topical similarities \cite{zhang2014}; Cai et al.\ \cite{Cai11} combine the result of LDA and translation models; Vasiljevi\'{c} et al.\ \cite{Vasiljevic16} explore combining a document's word count and topic model similarity into one measure; and Yang et al.\ \cite{Yang2013} form a generative probabilistic method to jointly model QA topic distributions and user expertise. In all of the above-mentioned topic modelling approaches similarity is calculated using the questions that exist in the database.

This work explores the possibility of deriving topic distributions from existing questions and answers, and proposes a method to relate these to the topic distribution of a novel question. Some work has been done in this direction; Zolaktaf et al.\ \cite{Zolaktaf11} model the question topics and then use them to condition the answer topics. This work proposes to model question and question-answer topics independently and then to learn a mapping between them. Furthermore, it extends topic modeling to include distributed word representations.

\section{Methodology}
\label{sec:methodolgy}

A corpora $C$ of size $L=|C|$ consists of many question-answer pairs: $C=\{(q_1,a_1),(q_2,a_2), \dots, (q_L,a_L) \}$, where $Q=\{q_1,q_2,\dots,q_L\}$ and $A=\{a_1,a_2, \dots, \linebreak a_L\}$, $\forall(q_i,a_i) \in C:q_i \in Q, a_i \in A$,  are question and answer sets (respectively). 

Questions in a CQA corpora tend to be shorter than answers and may contain few relevant words, which limits a model's ability to discover underlying trends. An approach to mitigate this is to assume that each question $q_i$ contains its text, and possibly keywords, a title, and a description. This assumption is not a requirement as meta-data may not always be available; however its absence may limit the ability of a model to represent the questions. We discuss this further in this section and propose methods to overcome the problem of question sparsity. Furthermore, each question in a QA corpora may have multiple answers and these are concatenated to form each element $a_i$ as they all provide contextual information that can be exploited to determine the question's relevance.

\begin{figure}[t]
	\centering
	\includegraphics[width=.92\linewidth]{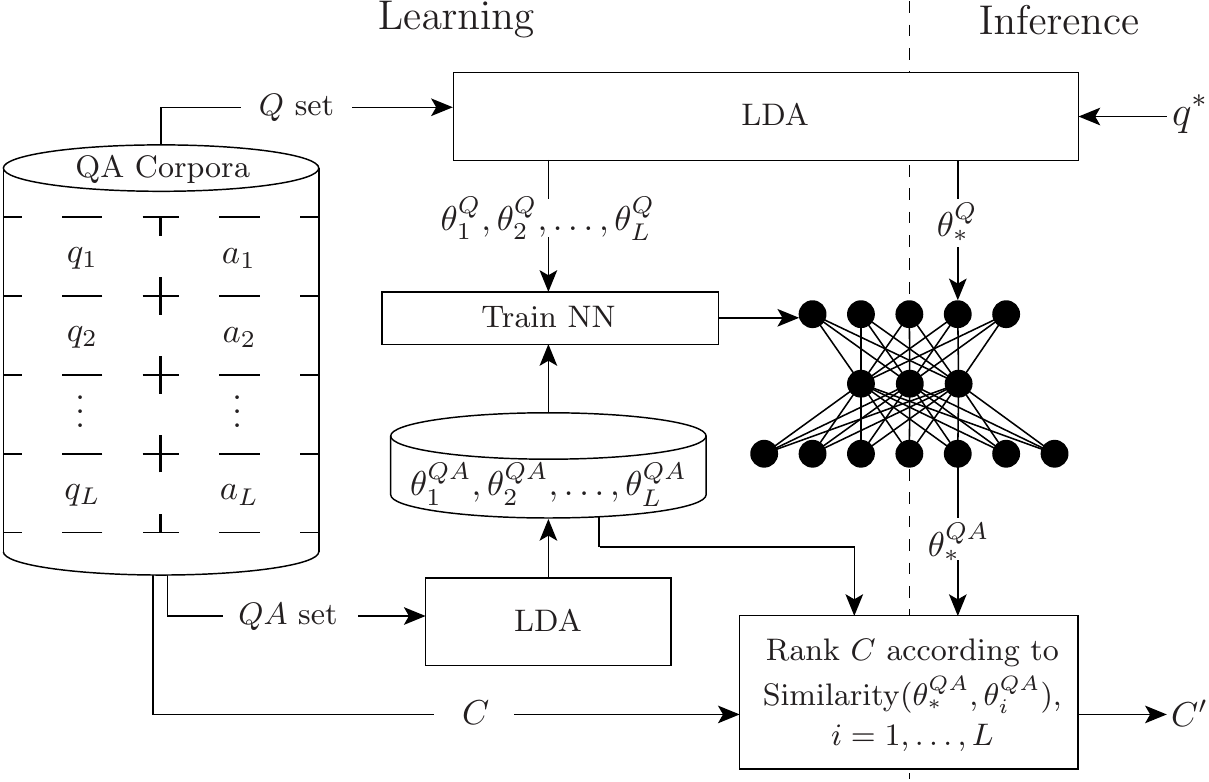}
	\caption{An overview of the system's training and inference (see text for details).}\label{fig:method}
\end{figure}

Figure \ref{fig:method} presents the proposed methodology. The task of similar question retrieval implies ranking the pairs contained in the QA Corpora ($C$) according to their similarity to a query question $q^*$, producing a partially ordered set $C'$ such that its first element has the highest similarity (the top, say, ten elements of which can then be returned as suggestions). In the learning phase of the proposed methodology, the QA corpora is used to train two topic models (Section \ref{subsec:LDADWR}): LDA on the set $Q$, and LDA on the set $QA$, in which each pair $(q_i,a_i)\in C$ is concatenated to form a single document. This results in topic distributions associated with the sets $Q$ and $QA$ and each element contained therein ($\theta^{Q}_i$ and $\theta^{QA}_i$ respectively). A regression model is trained using the samples $\theta^{Q}_i$ and $\theta^{QA}_i$ (Train NN) to learn the translation function between the Q and QA topic distributions  (Section \ref{subsec:NMR}).  During inference the $Q$ set LDA model is used to determine the topic distribution of a query question ($\theta^{Q}_*$) which is translated to a QA topic distribution ($\theta^{QA}_*$) using the regression model. Finally, a similarity measure (Section \ref{subsec:NMR}) is used to rank the QA Corpora ($QA$) according the similarity between each pair's topic distribution ($\theta^{QA}_i$)  and the query question's QA topic distribution ($\theta^{QA}_*$) obtained from the regression model. The LDA and regression models are discussed in more detail in the following subsections.

\subsection{Latent Dirichlet Allocation \& Distributed Word Representation}
\label{subsec:LDADWR}

In this work we assert that topic modeling provides a representation of the elements in $C$ that facilitates the discovery of semantically similar questions; particularly when these similar questions do not have words in common.

\begin{figure}[t]
	\centering
	\includegraphics[width=.8\linewidth]{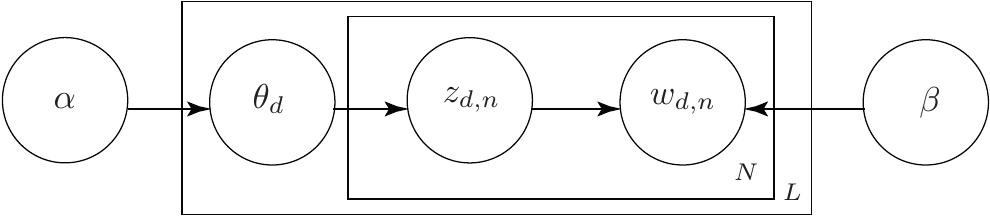}
	\caption{LDA plate notation, $\beta$ is the Dirichlet prior on the per-topic word distribution.}
	\label{fig:LDA}
\end{figure}

Latent Dirichlet Allocation (LDA) \cite{Blei2003} is a generative probabilistic model that enables us to describe a collection of discrete observations in terms of latent variables. The plate notation representing LDA is presented in Figure \ref{fig:LDA}.
When applied to a corpora, LDA models the generation of each document by means of two stochastic independent processes and can be summarised as follows
\begin{enumerate}[noitemsep,topsep=0pt,leftmargin=*]
 \item For each document $d$ in the collection $D$, randomly choose a distribution over topics $\theta_{d} \sim Dir(\alpha)$, where $\alpha$ is the Dirichlet prior.
 \item For each word $w_n$ in document $d$:
 \begin{enumerate}[noitemsep,topsep=0pt]
 \item choose a topic from the distribution over topics in Step 1.\ $z_{d,n}\sim \Multi(\theta_{d})$;
 \item choose a word from the vocabulary distribution $w_{d,n} \sim \Multi(\phi_{z_{d,n}})$.
 \end{enumerate}
\end{enumerate}
After learning a corpora's latent variables a topic is represented as a multinomial distribution of words, and a document by a multinomial distribution of topics.

The LDA algorithm described above treats words as explicit constraints, which inhibits its effectiveness when words are rare. A solution is to treat words as features \cite{Petterson2010} and the method used to calculate a word's features then influences its topic membership. This allows us to exploit a word's semantic similarity to augment information in short questions by giving similar topic membership probabilities to semantically equivalent words. For example, the words ``educator'', ``education'', ``educational'', and ``instruction'' should have similar probabilities within a certain topic, even if some of these words appear rarely in the corpus. 

Mikolov et al.\ \cite{Mikolov2013} introduced the continuous bag-of-words and Skip-gram neural network models that produce a continuous-valued vectorial word representation by exploiting the content of large textual databases. Distances between these vectors are proportional to the semantic difference of the words they represent, and thus these vectors can be used as features in many NLP tasks. In this work, the Word2vec vector representation is used to group semantically related words; its use for this application was first proposed by Petterson et al.\ \cite{Petterson2010}. 

In the original LDA algorithm, a word is generated by the process $w_{d,n} \sim \Multi(\phi_{z_{d,n}})$ where $\phi_{z_{d,n}}$ is the multinomial distribution ($\Multi$) of word probabilities in topic $z_{d,n}$ over the whole vocabulary. In order to introduce the distributed representation of words, we define a function $v : \mathbb{R} \to \mathbb{R}^r$ that maps a word to its vectorial representation learnt by Word2vec, where $r$ the number of latent features used for the distributed word representation (in practice this function is represented by a matrix $\omega \in \mathbb{R}^{N\times r}$, where $N$ is the vocabulary size). Given two words $w$ and $w'$  their semantic similarity can be found by applying the cosine similarity function, see Eq.\ \eqref{eq:cossim}, to their vectorial representations, i.e.\ $\similarity(v(w),v(w'))$. A set of words that are similar to $w$, $\Omega_w$, can be obtained by defining a threshold $\tau$ such that $\Omega_w = \{ w'\,|\,\similarity(v(w),v(w'))> \tau\}$. This set can be used to define an alternative distribution of word probabilities $\phi'_{z_{d,n}}$ for topic $z_{d,n}$, in which the probability of a word $w$ is given by
\begin{equation}
\phi'_{z_{d,n}} (w)=\frac{1}{c} \sum\limits_{w' \in \Omega_w} \exp\left ( \phi_{z_{d,n}}(w') \similarity\left(v(w),v(w')\right)\right ),
\end{equation}
where $c$ is a normalisation factor. This modified distribution gives a high probability to semantically related words. Finally we consider each word $w$ to be sampled from a linear combination of the original and modified distributions
\begin{equation}
w_{d,n} \sim \lambda \Multi(\phi_{z_{d,n}}) + (1-\lambda)\Multi(\phi'_{z_{d,n}}), \quad 0 \leq \lambda \leq 1.
\end{equation}
We fixed $\lambda$ to $0.9$ so that the results of standard LDA are not excessively altered.

In order to implement this modification, the Gibbs Sampling-based algorithm proposed by \cite{Yao2009} was adapted so that at each step the probability of topic $t$ being present in document $d$ given word $w$ is estimated as follows:
\begin{align}\label{eq:gibbs}
p(z=t \mid w) = & \,\frac{\alpha \beta}{\beta V + n_{.\mid t}} + \frac{n_{t\mid d} \beta}{\beta V + n_{.\mid t}} + 
\frac{\left (\alpha + n_{t\mid d}\right )  \lambda n_{w\mid t}}{\beta V + n_{.\mid t}} \nonumber\\
&+\frac{1-\lambda}{c}\sum\limits_{w' \in  \Omega_w}\exp\left(\frac{n_{w'\mid t}\similarity\left(v(w),v(w')\right)}{n_{.\mid t}}\right),
\end{align}
where $n_{w\mid t}$ is the number of words $w$ assigned to topic $t$, $n_{t\mid d}$ is the total number of words in document $d$ assigned to topic $t$, $n_{.\mid t} = \sum_{w}n_{w\mid t}$, $\alpha=35/T$ is the Dirichlet prior of the per document topic distribution (for number of topics $T$),  and $\beta=0.01$ \cite{Griffiths04,Yao2009}. Small values of $\alpha$ and $\beta$ result in a fine-grained decomposition into topics that address specific areas \cite{Griffiths04,Yao2009}.

This method is applied to two document collections (Figure \ref{fig:method}), $Q$ and $QA$, which results in two topic models: $T_Q=\{ 1,\dots,K_Q \}$ in which each question $q_i$ is represented by the distribution of topics $\theta_i^Q$; and $T_{QA}=\{ 1,\dots,K_{QA} \}$ in which each pair $(q_i,a_i)$ is represented by the distribution of topics $\theta_i^{QA}$.

\subsection{Nonlinear Multinomial Regression} 
\label{subsec:NMR}

When a query question $q^*$ is entered the left-to-right method \cite{Wallach09} is used to infer its topic distribution, $\theta^{Q}_*$. A regression model is therefore needed to obtain an estimate of $\theta^Q_*$ mapped to a distribution of topics in the $QA$ set, $\theta^{QA}_*$. Mapping the distribution of question topics to the distribution of question-answer topics avoids problems that occur when limited vocabularies are used in a question. This information is augmented with that derived from the set of answer terms, thus by mapping a query question to the space of question-answers it is possible to calculate its similarity using words that do not exist in the question vocabulary (and therefore are not represented in the topic distribution $T_Q$). Performing this mapping also provides a means to model the relationship between question semantics and existing question-answer semantics (which will be discussed further in Section \ref{sec:discussion}): given a query question $q^{*}$ the model estimates a topic distribution in the space of concatenated questions and answers, which can be compared to the distributions of existing QA pairs.

Determining the topic distribution in the space of documents comprising questions and answers, given the topic distribution of a new question is a problem of multinomial regression. For which we use a multilayer perceptron neural network (NN), which are nonlinear multinomial regression models \cite{Ripley96, Bentz2000}. The NN is trained using the set of topic distributions for each document in $Q$ and $QA$, $\theta^{Q}_i$ and $\theta^{QA}_i$ (respectively) where $i = 1,\dots,L$, and therefore the input and output layers have as many nodes as the number of topics used to model these sets, $K_Q$ and $K_{QA}$ (respectively). Sigmoid activation functions are used in the hidden layer and softmax in the output layer to ensure that outputs sum to one.

In application the input of the NN is the topic distribution of the query question according to latent topic model of the existing questions, represented by $\theta^{Q}_*$, and its output is an estimate of its distribution in the QA latent topic model, $\theta^{QA}_*$. The cosine similarity measure allows us to rank existing questions $q_i$ according to their similarity to $\theta^{QA}_*$, i.e.
\begin{equation}
\similarity\left(\theta^{QA}_*, \theta^{QA}_i\right)= \frac{\theta^{QA}_* \cdot \theta^{QA}_i}{\lVert \theta^{QA}_* \rVert \lVert \theta^{QA}_i \rVert},
\label{eq:cossim}
\end{equation}
where $\lVert x \rVert$ is the length of vector $x$, and therefore the most similar existing questions appear at the top of the ranked list that is output by the system.

\section{Evaluation}
\label{sec:experiment}

This section describes the data, experimental setup, and comparison algorithms used to evaluate the proposed approach.

\subsection{Data}
Four categories, derived from two different CQA sources, are used for the evaluation. The first two are the \textit{Health} and \textit{Computers \& Internet} (referred to herein as Computers) categories in the publicly available Yahoo!\ Questions L6 (Yahoo!\ Answers Comprehensive Questions and Answers version 1.0) dataset\footnote{Available from \url{http://webscope.sandbox.yahoo.com/catalog.php?datatype=l}}. The second two are the \textit{Physics} and \textit{Geographic Information Systems} (GIS) categories taken from the publicly available StackExchange (SE) dataset\footnote{Available from \url{https://archive.org/details/stackexchange}}. The question sets extracted from the Yahoo! dataset were created by concatenating the question text and description (when available), and the question sets extracted from the SE dataset were created by concatenating the question title, tags, and text. The answer sets were created by concatenating all the answers provided by different users for a particular question. Table \ref{tab:datasets} summarises these datasets. 

Preprocessing was performed before data use: stop words were removed using Mallet's standard English list ($543$ words), non-English characters were removed, and lemmatization was performed to reduce the number of inflected word forms.

Fifty randomly selected questions from each category were used for testing and the remaining pairs were used as training data. Therefore four models were calculated using each algorithm, one for each category. The output of each model (the top ten most similar results for each test question) were manually labelled as relevant or not and this was used to calculate the evaluation statistics.

\begin{table*}[!t]
\caption{Summary of the datasets}
\label{tab:datasets}
\begin{center}
\begin{scriptsize}
\begin{tabular}{lcccc}
\toprule
 & Y! Health & Y! Computers & SE Physics & SE GIS \\
\midrule
\# of Questions & \num{40050}  & \num{40050} & \num{41201} & \num{36520}  \\
\# of Answers & \num{133747}  & \num{116946} & \num{77767} & \num{59873}  \\
\# of Answers per Question & \num{3.34} & \num{2.92} & \num{1.89} & \num{1.64}   \\
Average Question Length (\# of words) & \num{22.26} & \num{27.33} & \num{122.31} & \num{145.31}   \\
Average Answer Length (\# of words) & \num{27.33} & \num{67.91} & \num{195.77} & \num{108.32}   \\
\bottomrule
\end{tabular}
\end{scriptsize}
\end{center}
\end{table*}

The Word2vec model requires training in order to learn the word embedding space, and this was realised using an additional corpus of Google news and Yahoo! Questions QA pairs (from categories other than those presented previously). The reason for including documents form Yahoo! Questions in this corpus is that it enables words that are specific to the dataset---such as abbreviations, misspellings, and technical jargon---to be learnt.

A modified version of Mallet, which implements the Gibbs sampling method proposed by Yao et al.\ \cite{Yao2009}, was used for Topic Modeling. The number of topics were empirically set to $140$ and $160$ for the $Q$ and $QA$ sets (respectively) and the size of the neural network's hidden layer was empirically set to $180$ using $100$ questions-answer pairs (these were subsequently removed from the corpus).

\subsection{Results}

The proposed method, referred to henceforth as LDA$^+$, was compared to four state-of-the-art algorithms: Translation1, the IBM translation approach proposed by Jeon et al.\ \cite{Jeon2005}; Translation2, the combined translation and query-likelihood language model proposed by Xue et al.\ \cite{xue2008}; an autoencoder based method proposed by Socher et al.\ \cite{Socher11}; to establish the benefit of word2vec, LDA$^*$ (as described within Section \ref{sec:methodolgy} excluding word2vec); and to establish the benefit of the regression stage, LDA$^\dagger$ (as described within Section \ref{sec:methodolgy} excluding the regression step).

Mean Average Precision (MAP) and Precision at $N$ (P@$N$) are used to summarise retrieval performance within each category. The autoencoder was found to be computationally infeasible when applied to the described datasets and therefore its retrieval performance is not presented. The results obtained using the remaining methods are presented in Table \ref{tab:overall_results}. A cursory validation of these results was performed by comparing the translation methods' figures to those presented in the literature using the same method and data source (but not the same partitioning) and they fall within the observed range \cite{Cai11,Zhou2013, singh2012, Zhou2013, Lee2008}.

The results show that in all of the datasets, LDA$^+$ outperforms all other methods. However, the difference is much more pronounced when the length of the question and answers increase (as is the case in the SE datasets). In this situation, the translation methods fail to find relevant documents whereas all of the LDA methods do (due to the increase in information). It is difficult to separate the performances of LDA with Word2vec and LDA with regression (LDA$^\dagger$ and LDA$^*$), but when combined (LDA$^+$) a performance increase is observed.

\begin{table*}[!t]
\caption{Performance of each method using two Yahoo! Questions categories and two StackExchange categories (to two significant figures). The highest results for each measure and category are in bold and italics indicate statistical significance when compared to LDA$^+$ using a paired two-sample t-test with an alpha level of $0.05$.}
\label{tab:overall_results}
\begin{center}
\begin{scriptsize}
\begin{tabular}{llcccccc}
\toprule
Method & Category & MAP & P@1 & P@2 & P@4 & P@7 & P@10 \\
\midrule
Translation1 & \multirow{5}{*}{ Y! Health} & 0.38 & 0.64 & 0.60 & 0.51 & 0.53 & 0.51 \\
Translation2 &  & 0.36 & 0.50 & 0.52 & 0.49 & 0.48 & 0.45 \\
LDA$^\dagger$  & & \textit{0.28} & 0.50 & 0.52 & 0.45 & \textit{0.42} & \textit{0.38} \\
LDA$^*$ & & \textit{0.29} & \textit{0.48} & \textit{0.51} & {0.48} & \textit{0.44} & \textit{0.41} \\
LDA$^+$ & & \textbf{0.43} & \textbf{0.66} & \textbf{0.64} & \textbf{0.57} & \textbf{0.55} & \textbf{0.53} \\
\midrule
Translation1 & \multirow{5}{*}{ Y! Computers} & \textit{0.21} & 0.50 & 0.41 & 0.36 & \textit{0.33} & \textit{0.31} \\
Translation2 & & 0.28 & 0.46 & 0.44 & 0.42 & 0.38 & 0.38  \\
LDA$^\dagger$  & & 0.26 & 0.50 & 0.43 & 0.39 & 0.37 & 0.34 \\
LDA$^*$ & & 0.31 & 0.50 & 0.48 & 0.45 & 0.43 & 0.394 \\
LDA$^+$ &  & \textbf{0.35} & \textbf{0.52} & \textbf{0.55} & \textbf{0.49} & \textbf{0.48} & \textbf{0.46} \\
\midrule
Translation1 & \multirow{5}{*}{ SE Physics} & \textit{0.34} & \textit{0.36} & \textit{0.41} & \textit{0.45} & \textit{0.43} & \textit{0.42} \\
Translation2 & & \textit{0.30} &  \textit{0.38} & \textit{0.35} & \textit{0.40} & \textit{0.40} & \textit{0.39} \\
LDA$^\dagger$    & & 0.67 & 0.82 & 0.78 & 0.76 & 0.75 & 0.74 \\
LDA$^*$  & & \textit{0.62} & 0.72 & \textit{0.78} & 0.80 & 0.75 & \textit{0.71} \\
LDA$^+$  &  & \textbf{0.71} &  \textbf{0.86} & \textbf{0.89} & \textbf{0.84} & \textbf{0.79}  & \textbf{0.77} \\
\midrule
Translation1 & \multirow{5}{*}{ SE GIS} & \textit{0.19} & \textit{0.32} & \textit{0.30} & \textit{0.31} & \textit{0.33} & \textit{0.30} \\
Translation2 & & \textit{0.14} & \textit{0.30} & \textit{0.27} & \textit{0.27} & \textit{0.23} & \textit{0.21} \\
LDA$^\dagger$ & & \textit{0.59} &  \textit{0.76} & \textit{0.77} & \textit{0.73} & \textit{0.70} & \textit{0.69} \\
LDA$^*$ & & \textit{0.59} & 0.92 & 0.86 & \textit{0.77} & \textit{0.70} & \textit{0.66} \\
LDA$^+$ &  & \textbf{0.75} &  \textbf{0.98} & \textbf{0.92} & \textbf{0.85} & \textbf{0.83} & \textbf{0.82} \\
\bottomrule
\end{tabular}
\end{scriptsize}
\end{center}
\end{table*}

\section{Discussion}
\label{sec:discussion}

Within the translation based approaches \cite{Jeon2005,xue2008} the translation probabilities of equal source and target words are fixed to $1$. This forces questions that share words in common with the query question to be highly ranked. Conversely, LDA$^\dagger$, LDA$^*$, and LDA$^+$ perform matching based upon shared topics, and inherently accounts for words that represent multiple concepts by decreasing their probabilities in the topics that they appear. To illustrate this, Table \ref{tab:examples} presents an example of retrieved questions using LDA$^+$ and the two translation based approaches\footnote{Mistakes in the questions are original to the data} (the points discussed in this section were observed in all of the categories but to save space we present examples from the Health category). In the first example, presented in the top half of the table, the QA pairs retrieved by LDA$^+$ do not contain the words ``lift'' and ``weight'' even though they are relevant to the query. The excessive contribution from the word ``weight'' causes the translation models to retrieve questions that are related to body weight instead of weight lifting. The second example illustrates a query in which all the retrieved QA pairs are relevant. As before, the translation methods result in questions that have words in common with the query question (as does LDA$^+$); in this case Translation2 associates a high translation probability between ``hair'' and ``mustache'' (sic).

\begin{table*}[!t]
\caption{Examples of the top two retrieved QA pairs for each method given a query question (using the Yahoo! Health category).}
\label{tab:examples}
\begin{center}
\begin{scriptsize}
\begin{tabular}{p{0.24\linewidth}p{0.24\linewidth}p{0.24\linewidth}p{0.24\linewidth}}
\toprule
Query Question & & Retrieved QA Pairs \\
\cmidrule{2-4}
 & LDA$^+$ & Translation1 & Translation2 \\
 \midrule
How many days a week should you lift weight? & When are you to old to build muscle mass from work out? & How do I gain body weight? & My weight is 90lb how could I gain more weight? \\
\rule{0pt}{3ex}
& What is a 1 set rep? & My weight is 90lb how could I gain more weight? & Will lose weight faster than average if I workout? \\
\midrule
Can you make your hair grow faster & Is there any any way to get rid of razor bump? & What important function do our body hair play? & How can I make my mustache grow faster? \\
\rule{0pt}{3ex}
& What's the best herbal remindie for hair loss?  & What can be do to prevent hair loss? & What is hair? \\
\bottomrule
\end{tabular}
\end{scriptsize}
\end{center}
\end{table*}

Table \ref{tab:topics} demonstrates the benefit of performing the multinomial regression. It presents the representative words (those that have high probability in the topic's word distribution) of three of the topics derived from the question ($Q$) set and the question+answers ($QA$) set. It demonstrates that the topics derived from the $QA$ set better represent the themes that appear in health documents, whilst the topics of derived from the $Q$ set are less distinguishable. For example, the words in Topic 3 appear to represent depression, however, the words derived from the $QA$ set are more coherent. This is because of the limited vocabulary used in questions and their typically short length.

Furthermore, the topics derived from the $Q$ set tend to represent the semantics of expressions commonly used in questions (and not in answers), for example the phrases ``an effective method'' and ``effective treatment''. The word ``effective'' in the topics derived from the $QA$ set is associated with the topic representing medical products. Consequently, when a question such as ``What is an effective sleeping aid?'' is posed to a model trained on the $QA$ set, topics in which the words ``method'' and ``treatment'' have high probability would not be considered. The model trained on the $Q$ set, however,  results in a high probability of Topic 1, and the regression stage of LDA$^+$ causes this to be mapped to the distribution in which the words ``treatment'' and ``method'' have higher probabilities. Another example is provided by Topic 2, here the word ``result'' is often mentioned in questions posed by those who have performed medical tests, while in answers the word usually refers to the results of health research studies.

\begin{table*}[!t]
\caption{Words that comprise the topics derived from the questions ($Q$) and question-answer ($QA$) sets of the Yahoo! Health category.}
\label{tab:topics}
\begin{center}
\begin{scriptsize}
\begin{tabular}{cp{0.92\linewidth}}
\toprule
Topic & $Q$\\
\midrule
1 & treatment, effective, method, suggest, special, option, undergo, acupuncture, indian, prognosis, bad, acupuncture\\
2 & test, result, show, urine, blood, negative, positive, pap, pass, testing, smear, screen, lab, tuberculosis, perform \\
3 & depression, suffer, depress, deal, solution, seek, advise, cost, remain, viagra, clinical, dysfunction, overcome, admit \\
\midrule
Topic & $QA$\\
\midrule
1 & pill, product, work, market, effective, company, fda, safe, ingredient, call, wont, sell, approve, brand, generic \\
2 & study, research, show, find, percent, report, health, american, accord, result, evidence, national, researcher \\
3 & depression, depress, feel, mood, medication, talk, anxiety, anti, therapy, psychiatrist, antidepressant \\
\bottomrule
\end{tabular}
\end{scriptsize}
\end{center}
\end{table*}

\section{Conclusions}\label{sec:conclusions}

This paper has presented a novel model that fuses topic modelling with Word2vec and a regression stage for ranking relevant questions-answer pairs within Collaborative Question Answering platforms. The performance of the proposed method has been evaluated using several real-world datasets, and it has been shown to outperform translation based methods and LDA with each innovation separately in all cases. Most notably when the dataset contains long questions and answers. It achieves this by allowing the model to overcome the differences in vocabulary used in questions and answers, helps to deal with the sparsity often encountered in questions (due to their relatively short length), and allows the method to exploit all available information.


\end{document}